\magnification=\magstep1   
\tolerance 10000
\baselineskip=12pt
\def\real{{\tt I\kern-.2em{R}}}
\def\nat{{\tt I\kern-.2em{N}}}
\def\hyper#1{\ ^*\kern-.2em{#1}}
\def\r#1{{\rm #1}}
\def\b#1{{\bf #1}}
\def\power#1{{{\cal P}(#1)}}
\def\Hyper#1{\hyper {\eskip #1}}
\def\eskip{\hskip.25em\relax}
\def\hypernat{{^*{\nat }}}

\def\st#1{{\tt st}(#1)}
\def\qed{{\vrule height6pt width3pt depth2pt}\par\medskip}
\hoffset=.25in
\hsize 6.00 true in
\vsize 8.85 true in
\font\eightrm=cmr9
\def\m@th{\mathsurround=0pt}

\def\id{\par\hangindent2\parindent\textindent}
\def\textindent#1{\indent\llap{#1}}
\centerline{\bf Probability Models and Ultralogics}\par\bigskip 
\centerline{Robert A. Herrmann}\par\medskip
\centerline{Mathematics Department}
\centerline{U. S. Naval Academy}
\centerline{572C Holloway Rd.}
\centerline{Annapolis, MD 21402-5002}
\centerline{7 DEC 2001, Last Revision 20 MAY 2013.}\bigskip
{\leftskip=0.5in \rightskip=0.5in \noindent {\eightrm {\it Abstract:} In this paper, we show how nonstandard consequence operators, ultralogics, can generate the general informational content displayed by probability models. In particular, a model that states a specific probability that an event will occur and those models that use a specific distribution to predict that an event will occur. These results have many diverse applications and even apply to the collapse of the wave function.\par}}\par\bigskip
 \noindent {\bf 1. Introduction.}\par\medskip  In [1], the theory of nonstandard consequence operators is introduced. Consequence operators, as an informal theory for logical deduction, were introduce by Tarski [2]. There are two such operators investigated, the {\it finite} and the {\it general} consequence operator. Let $\r L$ be any nonempty set that represents a language and ${\cal P}$ be the set-theoretic power set operator.\par

\medskip
{\bf Definition 1.1.} A mapping $\r C\colon \power{\r L} \to \power {\r L}$ is a {\it general} consequence operator (or closure operator) if for each $\r X,\ \r Y \in 
\power {\r L}$\par\smallskip

\indent\indent (i) $\rm X \subset C(X) = C(C(X)) \subset L$ and if\par\smallskip

\indent\indent (ii) $\rm X \subset Y$, then $\rm C(X) \subset C(Y).$\par\medskip

\noindent A consequence operator C defined on L is said to be {\it finite} ({\it finitary}, or {\it algebraic}) if it satisfies\par\smallskip

\indent\indent (iii) $\rm C(X) = \bigcup\{C(A)\mid A \in F(\r X)\},$ where $\r F$ is the finite power set operator.\par

\medskip
{\bf Remark 1.1.} The above axioms (i) (ii) (iii) are not independent. Indeed, 
(i) (iii) imply (ii).\par\medskip

In [1], the language $\r L$ and the set of all consequence operators defined on $\r L$ are encoded  and embedded into a standard superstructure ${\cal M} = \langle {\cal N}, \in, = \rangle.$ This standard superstructure  
is further embedded into a nonstandard and elementary extension $\Hyper {\cal M}= \langle \Hyper {\cal N},\in,=\rangle.$  For convince, $\Hyper {\cal M}$ is considered to be a $2^{\vert \cal M\vert}$-saturated enlargement. Then, in the usual constructive manner, $\Hyper {\cal M}$ is further embedded into the superstructure, the Grundlegend structure, ${\cal Y} = \langle Y, \in ,= \rangle$ where, usually, the nonstandard analysis occurs. In all that follows in this article, the Grundlegend superstructure ${\cal Y}$ is altered by adjoining to the construction of ${\cal M}$ a set of atoms that corresponds to the real numbers. This yields a $2^{\vert \cal M\vert}$-saturated enlargement $\Hyper {\cal M}_1$ and the corresponding Extended  Grundlegend structure ${\cal Y}_1$ [3].\par

\vfil\eject
\noindent {\bf 2. The Main Result.}\par \medskip 
To indicate the intuitive ordering of any sequence of events, the set $\r T$ of Kleene styled``tick'' marks, with a spacing symbol, is used [4,  p. 202] as they might be metamathematically abbreviated by symbols for the non-zero natural numbers.  
Let $\r G \in \r L_1$ be considered as a fixed description for a source that yields, through application of natural laws or processes, the occurrence of an event described by $\r E \in \r L_1.$ Further, the statement $\r E' \in \r L_1$ indicates that the event described within the 
statement $\r E$ did not occur. Let $\r L = \{\r G\} \cup \{\r E, \r E'\} \cup \r T.$ As usual, ${\rm G,\ E,\ E'}$ are assumed to contain associated encoded general information. Note that for subsets or members of $\r L$ bold notation, such as $\b G$, denotes the image of $\r G$ as it is embedded into ${\cal M}_1.$ \par
\medskip
{\bf Theorem 2.1.} {\it For the language ${\rm L}$ and any $p\in \real$ such that $0\leq p\leq 1,$ where $p$ represents a Bernoulli trials probability that an event will occur, there exists an ultralogic $ P_{p}$ with the following properties.\par

$1.$ When $P_{p}$ is applied to $\Hyper {\{\b G \}} = \{\b G \}$ a hyperfinite sequence of labeled event statements $\b E$ or $\b E'$ is obtained that explicitly generates the sequence $\{a_1,\ldots,a_n,\ldots,\kern-.3em\hyper a_{\nu} \}$. 
For any ``n'' trials, the hyperfinite sequence $\{a_1,\ldots,a_n,\ldots,\kern-.3em\hyper a_{\nu} \}$ yields a finite ``event'' sequence $\{a_1,\cdots,a_n\}.$ Further, for each nonzero natural number $j$ each $a_j$ is the cumulative number of successes $\b E$ for ``j'' trials. These sequences mimic the behavior of the cumulative successes $\b E$ for Bernoulli trials without introducing specific Bernoulli trial requirements. 
\par
 
 $2.$ The events $\b E$ in $1$ determine a sequence $g_{ap}$  of relative frequencies that converges to $p,$ where $g_{ap}(n) = (n,a(n))= a(n)/n$.\par

$3.$ The sequence of relative frequencies $g_{ap}$ is what one would obtain from Bernoulli trial required random behavior.}\par

\medskip
{\bf Proof.} 
 All of the objects discussed will be members of an informal set-theoretic structure and slightly abbreviated definitions, as also discussed in [3, p. 23, 30-31], are utilized. [Indeed, all that is needed is an intuitive superstructure.] As usual $\nat$ is the set of all natural numbers including zero, and $\nat^{> 0}$ the set of all non-zero natural numbers.\par

Let $A=\{a\mid (a\colon \nat^{>0} \to \nat)\land(\forall n(n \in \nat^{>0} \to (a(1)\leq 1\ \land\  0\leq a(n+1)-a(n)\leq 1)))\}.$ Note that the special sequences in $A$ are non-decreasing and for each $n \in \nat^{>0},\ a(n)\leq n.$ Obviously $A \not=\emptyset,$ for   
the basic example to be used below, consider the sequence $a(1) =0, \ a(2) =1,\ a(3) =1, \ a(4) =2,\ a(5) =2,\ a(6) = 3,\ a(7)= 3, \ a(8) = 4, \ldots$ which is a member of $A.$ Next consider the must basic representation $Q$ for the non-negative rational numbers
where we do not consider them as equivalence classes.   Thus 
$Q = \{(n,m)\mid (m \in \nat)\land (n \in \nat^{>0})\}.$\par  

For each member of $A,$ consider the sequence $g_a \colon \nat \to Q$ defined by $g_a(n) = (n,a(n)).$ Let $F$ be the set of all such $g_a$ as $a \in A.$ Consider from the above hypotheses, any 
$p \in \real$ such that $0\leq p \leq 1.$ We show that for any such $p$ there exists an $a \in A$ and a $g_{ap} \in F$ such that $\lim_{n \to \infty}g_{ap}(n) = p.$ For each $n\in \nat^{> 0},$ consider $n$ subdivision of $[0,1],$ and the corresponding intervals $[c_k,c_{k+1}),$ where $c_{k+1} - c_k =1/n,\ 0\leq k < n,$ and $ c_0=0,\ c_n = 1.$ If $p=0,$ let $a(n) = 0$ for each $n \in \nat^{> 0}.$  Otherwise, using the customary covering argument relative to such intervals, the number $p$ is a member of one and only one of these intervals, for each $n \in \nat^{> 0}.$ Hence for each such $n>0,$ select the end point $c_k$ of the unique interval $[c_k,c_{k+1})$ that contains $p.$ Notice that for $n = 1,$ $c_k = c_0 = 0.$ For each such selection, let $a(n) = k.$ Using this inductive styled definition for the sequence $a,$ it is immediate, from a simple induction proof, that $a \in A,\  g_{ap} \in F,$ and that $\lim_{n \to \infty}g_{ap}(n) = p.$ For the basic example $a$ above, this yields $\{(1,0),(2,1),(3,1),(4,2),(5,2),(6,3),(7,3),(8,4), \ldots \}$ and $g_{ap}$ converges to $1/2.$ Let nonempty $F_p \subset F$ be the set of all such $g_{ap}.$ Note that for the set $F_p,$ $p$ is fixed and $F_p$ contains each $g_{ap},$ as $a$ varies over $A,$ that satisfies the convergence and special form requirements. Thus, for $0\leq p \leq 1,$ $A$ is partitioned into subsets $A_p$ and a single set $A'$ such that each member of $A_p$ determines a $g_{ap} \in F_p.$ The elements of $A'$ are the members of $A$ that are not so characterized by such a $p.$ Let ${\cal A}$ denote this set of partitions.  \par

Let $B = \{f\mid \forall n \forall m(((n \in \nat^{>0})\land (m \in \nat)\land (m \leq n)) \to ((f\colon ([1,n] \times \{n\}) \times \{m\} \to \{0,1\})\land (\forall j(((j \in \nat^{>0}) \land (1 \leq j \leq n)) \to (\sum_{j =1}^n f(((j,n),n),m) =m)))))\}.$
The members of $B$ are determined, but not uniquely, by each $(n,m)$ such that $(n \in \nat^{>0})\land (m \in \nat)\land (m \leq n).$ Hence for each such $(n,m),$ let $f_{nm}\in B$ denote a member of $B$ that satisfies the conditions for a specific $(n,m).$\par 

For a given $p,$ by application of the axiom of choice, with respect to ${\cal A},$ there is an $a \in A_p$ and a $g_{ap}$ with the properties discussed above. Also there is a  sequence $f_{na(n)}$ of partial sequences such that, when $n >1,$ it follows that ($\dag$) 
$f_{na(n)}(j) = f_{(n-1)a(n-1)}(j)$ as $1\leq j \leq (n-1).$  Relative to the above example, consider the following: 
$$f_{1a(1)}(1)  =0,$$
$$f_{2a(2)}(1)  =0,\ f_{2a(2)}(2)  =1,$$
$$f_{3a(3)}(1)  =0,\ f_{3a(3)}(2)  =1,\ f_{3a(3)}(3)  =0,$$
$$f_{4a(4)}(1)  =0,\ f_{4a(4)}(2)  =1,\ f_{4a(4)}(3)  =0,\ f_{4a(4)}(4) =1,$$
$$ f_{5a(5)}(1)  =0,\ f_{5a(5)}(2)  =1,\ f_{5a(5)}(3)  =0,\ f_{5a(5)}(4) =1,\ f_{5a(5)}(5) = 0, \cdots$$\par

It is obvious how this unique sequence of partial sequences is obtained from any $a \in A.$ For each $a \in A,$ let $B_a =\{f_{nm}\mid \forall n(n \in \nat^{>0} \to m = a(n))\}.$ Let $B_a^{\dag} \subset B_a$ such that each $f_{nm} \in B_a^{\dag}$ satisfies the partial sequence requirement ($\dag$). For each $n \in \nat^{>0},$ let $Pf_{na(n)}\in B_a^{\dag}$ denote the unique partial sequence of $n$ terms generated by
an $a$ and the ($\dag$) requirement. In  general, as will be demonstrated below, it is the $Pf_{na(n)}$ that yields the set of consequence operators as they are defined on $\rm L.$ Consider an additional map $M$ from the set $PF = \{Pf_{na(n)}\mid a\in A\}$ of these partial sequences into our  descriptive language $\rm L$ for the source $\rm G$  and events $\r E,\r E'$ as they are now considered as labeled by the Kleene tick marks. For each $n \in \nat^{>0},$ and $1 \leq j \leq n,$ if $Pf_{na(n)}(j) = 0,$ then $M(Pf_{na(n)}(j))= \r E'$ (i.e. ${\rm E' = E}$ does not occur); if $Pf_{na(n)}(j) = 1,$ then $M(Pf_{na(n)}(j))= \r E$ (i.e. E does occur), as 
$1\leq j \leq n,$ where the partial sequence $j = 1,\cdots, n$ models the intuitive concept of an event sequence since each $\r E$ or $\r E'$ now contains the appropriate Kleene ``tick'' symbols or natural number symbols that are an abbreviation for this tick notation.\par 

Consider the set of axiomless consequence operators, each defined on $\r L,$ $\r H = \{\r C(\r X,\{\r G\})\mid \r X \subset \r L \},$ where if  $\r G \in \r Y,$ then
$\r C(\r X, \{\r G\})(\r Y) = \r Y \cup \r X;$ if $\r G \notin \r Y,$ then $\r C(\r X, \{\r G\})(\r Y) = \r Y.$ Then for each $a \in A_p,$ $n \in \nat^{>0}$ and respective $Pf_{na(n)}=P_{na(n)},$ there exists the set of consequence operators $C_{ap}=
\{\r C(\{M(P_{na(n)}(j))\},\{\r G\})\mid 1\leq j \leq n\} \subset \r H.$
Note that from [1, p. 5], $\r H$ is closed under the finite $\vee$ and the actual consequence operator is 
$\r C(\{M(P_{na(n)}(1))\} \cup\cdots \cup  \{M(P_{na(n)}(n))\}, \{\r G\}).$ Applying a realism relation $\r R$ (i.e. in general, $\r R(\r C(\{\r G \})) = \r C(\{\r G \}) -\{\r G \}$) to $\r C(\{M(P_{na(n)}(1))\} \cup \cdots \cup\{M(P_{na(n)}(n))\},\{\r G\})(\{\r G\})$ yields the actual labeled or identified event partial sequence $\{M(P_{na(n)}(1)), \ldots, M(P_{na(n)}(n))\}.$\par

Due to the set-theoretic notions used, one now imbeds the above intuitive results into the superstructure 
${\cal M}_1 = \langle {\cal R}, \in , =\rangle$ which is further embedded into the nonstandard structure $\hyper {\cal M}_1 = \langle \Hyper {\cal R}, \in , = \rangle$ [3]. Let $p\in \real$ be such that $0\leq p\leq 1,$ where $p$ represents a theory predicted (i.e. a priori) probability that an event will occur. Applying a choice function $C$ to $\cal A,$ there is some $a \in A_p$ such that $g_{ap} \to p.$ Thus $\Hyper C$ applied to  $\hyper {{\cal A}}$ yields $\hyper {a} \in \hyper {A_p}$ and $\hyper {g_{ap}} \in \hyper {F_p}.$ Let $\nu \in \hypernat$ be any infinite natural number. The hyperfinite sequence $\{ {a_1},\ldots,a_n,\ldots,\kern-.3em\hyper {a_\nu}\}$ exists and corresponds to $\{a_1,\ldots, a_n\}$ for any natural number $n \in \nat^{>0}.$ Also we know that $\st{(\mu,\hyper a(\mu)}= p$ for any infinite natural number $\mu.$ Thus there exists some 
internal hyperfinite $Pf_{\nu\kern-.3em\hyper {a(\nu)}} \in \hyper {PF}$ with the 
*-transferred properties mentioned above. Since $\Hyper {\b H}$ is closed under hyperfinite $\vee,$ there is a $P_p \in \Hyper {\b H}$ such that, after application of the relation $\Hyper {\b R},$ the result is the hyperfinite sequence $S=\{\kern-.3em\hyper {M}(P_{\nu\kern-.3em\hyper {a(\nu)}}(1)), \ldots,\kern-.3em\hyper {M}(P_{\nu\kern-.3em\hyper {a}(\nu)}(j)),\ldots, \kern-.3em\hyper {M}(P_{\nu\kern-.3em\hyper {a}(\nu)}(\nu))\}.$ Note that if $j\in \nat,$ then we have that $\Hyper {\b E} = \b E$ or $\Hyper {\b E'} = \b E'$ as the case may be.\par 

An extended standard mapping that restricts $S$ to internal subsets would restrict $S$ to  $\{\kern-.3em\hyper {M}(P_{\nu\kern-.3em\hyper {a(\nu)}}(1)), \ldots,\hyper {M}(P_{\nu\kern-.3em\hyper {a}(\nu)}(j))\},$ whenever $j \in \nat^{>0}.$ Such a restriction map models the restriction of $S$ to the natural-world in accordance with the general interpretation given for internal or finite standard objects [3, p. 98]. This completes the proof.   \qed\par
\noindent {\bf Remark 2.1.} Obviously, for Theorem 2.1, each $\rm E$ or $\rm E'$ exist separately. The conclusions may be viewed conditionally and as ordered responses. That is, based upon the source, if only a single or a few $\rm E$ or $\rm E'$ are obtained, one would conclude that these events are among sets such as $S$ and they correspond to the probability statement if the trials continued under the exact same conditions. Also note that for any language $\rm L'$, where $\rm T \subset L',\ \b G \in \b L'$ and for internal $Y \subset \Hyper {\b L'}$, if $\b G \in Y$ and $P_p$ is applied to $Y$, then using the realism relation the same results are obtained as those using the language L. Further, $\rm \{E,E'\}$ can be replaced with a nonempty set of descriptions $\rm E \cup E'$, where for the sets $\rm E$ and $\rm E'$ it can be that $\Hyper {\bf E} \not= \b E$ and $\Hyper {\bf E'} \not= \b E.'$ Changes such as these should be taken into account when other specific languages are considered. \par\medskip
  In a recent paper [5], it has been shown that general logic-systems and finitary consequence operators are equivalent notions. Throughout all of the mathematical results that deal with ultralogics, two ultralogic processes are tacitly applied whenever necessary. For a nonempty hyperfinite set $X$, there is an internal bijection $f$ defined on $[1,\nu],\ \nu \in \hypernat^{>0}$ and $f\colon [1,\nu] \to X.$ Such an $f$ is a {\it hyperfinite choice operator (function)}.  When useful, this function can also be considered as inducing a simple order on $X$ via the simple order of $[1,\nu].$ For any nonempty simply ordered finite standard set $Y$ of cardinality $n$, an induction proof shows that there exists an order preserving bijection $g\colon [1,n]\to Y$ such that $g(i)<g(j),\ i,j \in [1,n],\ i < j.$ Consequently, for any hyperfinite set $X$ with a simple order such an order preserving internal $f$ exists. This (internal) bijection is the {\it hyperfinite order preserving choice operator (function)}. These two operators are considered ultralogics since they model two of the most basic aspects for deductive thought. \par
For Theorem 2.1, the labeling of each $\rm E',\ E$ is only used to differentiate between the occurrences or non-occurrences of an event relative to the source generator $\rm G.$  Thus, $S$ can be considered as representing a hyperfinite choice operator. The maps that are obtained by restricting such hyperfinite operators relative to $S$ are standard and internal hyperfinite (indeed, finite) choice operators.\par
\par \medskip

\noindent {\bf 3. Distributions.}\par\medskip 
Prior to considering the statistical notion of a frequency (mass, density) function and the distribution it generates, there is need to consider a finite {\bf Cartesian product} consequence operator. Suppose that we have a finite set of consequence operators $\rm {\cal C} = \{C_1, \ldots, C_m\},$ where each is defined upon its own language ${\rm L_k}$. Define the operator $\rm \Pi C_m$ as follows: for any $\rm X \subset L_1 \times \cdots \times L_m$, using the projections $\rm pr_k$, consider the Cartesian product $\rm pr_1(X)  \times \cdots \times pr_m(X)$. Then 
$ \rm \Pi C_m (X) = C_1(pr_1(X))\times \cdots \times C_m(pr_m(\r X))$ is a consequence operator on ${\rm L_1}\times \cdots \times {\rm L_m}$ [5, Theorem 6.3]. If, at least one $\rm C_j$ is axiomless, then $ \rm \Pi C_m (X)$ is axiomless. If each $\rm C_k$ is a finite and axiomless consequence operator, then $\rm \Pi C_m$ is finite. All of these standard facts also hold within our nonstandard structure under *-transfer.\par 

A distribution's frequence function is always considered to be the probabilistic measure that determines the number of events that occur within a {\bf cell} or ``interval'' for  a specific decomposition of the events into various definable and disjoint  cells. There is a specific probability that a specific number of events will be contained in a specific cell and  each event must occur in one and only one cell and not occur in any other cell. \par

For each distribution over a specific set of cells, $I_k$, there is a specific probability $p_k$ that an event will occur in  cell $I_k$. Assuming that the distribution does indeed depict physical behavior, we will have a special collection of $g_{ap_k}$ sequences generated. For example, assume that we have three cells and the three probabilities $p_1 = 1/4,\ p_2 = 1/2, \ p_3 = 1/4$ that events will occupy each of these cells. Assume that the number of ``experiments'' is 6. Then the three partial sequence might appear as follows $$\cases{g_{ap_1}= \{(1,1),(2,1),(3,1),(4,2),(5,2), (6,2)\}&\cr
                    g_{ap_2}= \{(1,0),(2,1),(3,2),(4,2),(5,2),(6,3)\}&\cr
                    g_{ap_3}= \{(1,0),(2,0),(3,0),(4,0),(5,1),(6,1)\}&\cr}$$
Thus after six experiments have occurred, 2 events are in the first cell, 3 events are in the second cell, and only 1 event is in the third cell. Of course, as the number of experiments continues the first sequence will converge to 1/4, the second to 1/2 and the third to 1/4. Clearly, these required $g_{ap_i}$ properties can be formally generated and generalized to any finite number $m$ of cells.\par

Relative to each factor of the Cartesian product set, all of the standard aspects of Theorem 2.1 will hold. Further, these intuitive results are embedded into the above superstructure and further embedded into our nonstandard structure. Hence, assume that the languages $\rm L_k = L_1$ and that the standard factor consequence operator $\rm C_k$ used to create the product consequence operator 
is a $C_{ap_k}$ of Theorem 2.1. Under the nonstandard embedding, we would have that for each factor, there is a pure nonstandard consequence operator $P_{p_k} \in \Hyper {\bf H_k}$. Finally, consider the nonstandard product consequence operator $\Pi P_{p_m}.$ For $\Hyper {(\{\b G_1\} \times \cdots \times \{\b G_m\})}=\{\b G_1\} \times \cdots \times \{\b G_m\},\ \b G_i = \b G$, this nonstandard product consequence operator yields for any fixed experiment number $n$, an ordered m-tuple, where one and only one coordinate would have the statement $\b E$ and all other coordinates the $\b E'.$ It would be these m-tuples that guide the proper cell placement for each event and would satisfy the usual requirements of the distribution. Hence, the patterns produced by a specific frequency function for a specific distribution may be rationally assumed to be the result of  ultralogic processes.\par
 The specific information contained in each $\rm G_i$ and the corresponding $\rm E_i,\ E_i'$ employed in this article are very general in character. Although it would be unusual, for the above results, it is not necessary to assume that for each $i$, $\rm G_i = G,\ E_i = E,\ E_i' = E'.$ Let the language $\rm L_1 \supset L.$ Note that, whether for distributions or the results in section 2, the nonstandard product consequence operator $\Pi P_{p_m}$ when applied to any internal $A_i \subset \Hyper {\bf L_1}$ such that $\b G_i \in A_i,\ 1\leq i\leq m$, where ${\bf E_i,\ E_i'} \notin A,$ yields, after application of the general hyperrealism relation $\Hyper {\b R}$ applied to each coordinate, the same result as if the application was only made to $\{\b G_1\} \times \cdots \times \{\b G_m\}.$ For such cases, it may not be necessary to apply the realism relation when observations are being considered since such observations should differentiate between the source $\rm G$ and the events by various means. \par

From a physical viewpoint, it should be obvious that, in this model, what is ``observed'' is the effect of the single coordinate projection that yields the E or $\rm E'$. Further, what constitutes an ``experiment'' and how the $\rm E,\ E'$ are described must be carefully considered.    
\par\medskip
\noindent {\bf 4. Collapse of the Wave Function.}\medskip 
Within quantum measure theory, the notion of the Copenhagen interpretation that yields the collapse of the wave function is often criticized as an external metaphysical process [6]. However, this interpretation is consistent with the logic that models quantum measure theory. When a physical theory is applied to the behavior of a natural-system that actually alters such behavior, the theory can be represented by a axiomless finitary consequence operator $S^V_{N_i}$. By definition, $\Hyper {S^V_{N_i}}$ is an ultralogic.\par
As stated in [6, page 31,32] ``In other words, the wave function of the apparatus takes the form of a packet that is initially single but subsequently splits, as a result of the coupling to the system, into a multitude of mutually orthogonal packets, one for each value of $s$. Here the controversies over interpretation of quantum mechanics starts. . . . According to the Copenhagen interpretation of quantum mechanics, wherever a state vector attains the form of equation 5 [$\vert \Psi_1\rangle = \sum_s c_s\vert s\rangle\vert \Phi[s]\rangle$] it immediately collapses. The wave function, instead of consisting of a multitude of packets, reduces to a single packet, and the vector $\vert \Psi_1\rangle$ reduces to the corresponding element $\vert s\rangle\vert \Phi[s]\rangle$ of the superposition. To which element of the superposition it reduces one can not say. One instead assigns a probability distribution to the possible outcomes, with weights given by $w_s =\vert c_s \vert^2.$ '' \par
Applications of the process discussed in section 3 depend upon the types of ``cells'' being considered. The definition of ``cell'' is very general as the next application shows. Each cell can be but a single term within a finite or infinite series. If the ``multitude of mutually orthogonal packets'' is finite, then a finitary and axiomless $\Pi P_{p_m}$ applies immediately and yields the collapse. Significantly, $\Pi P_{p_m}$ eliminates all of the intermediate mathematical steps since $\Pi P_{p_m}$ relates any source specific information to any event specific information, where specific information generates the real physical content. \par
If the multitude of packets is an infinite set, then the Cartesian product notion would need to be defined in terms of ``mappings'' along with the axiom of choice. Since the internal $\Pi P_{p_m}$ exists for any $n \in \nat^{0>}$, then there exists such an operator $\Pi P_{p_\nu}$for any $\nu \in \hypernat^{0>}.$ This $\Pi P_{p_\nu}$ has all of the same first-order internal set-theoretic properties as each $\Pi P_{p_m}$.
In particular, when restricted to the standard infinite set of packets, application of the ultralogic $\Pi P_{p_\nu}$ yields the collapse. For both of these ultralogic collapse processes, the same remark 2.1 holds. \medskip 

\noindent {\bf 5. Additional Theorem 2.1 Information.}\par\medskip 

In this section, among other results, are presented the inductive processes that produce various cumulative event sequences and the Bernoulli-styled relative frequency sequences that are used in Theorem 2.1. That is, we look more closely at members of $A_p.$ As defined such a cumulative sequence $a
\colon \nat^{>0} \to \nat$ has this form when $a(1)= 0$ or $1,$ and $0\leq a(n+1) - a(n)\leq 1$ [or $a(n+1) = a(n)$ or $a(n+1) = a(n)+1$]. Hence, by trivial induction, for each $n \in \nat^{>0},$ $a(n) \leq n$ and if $a(1) = 0,$ $a(n)<n.$  A sequence is of this type if and only if it is a member of the set $A$ of Theorem 2.1. \par\smallskip
Let $p \in [0,1].$ If $p = 0,$ then for each $n \in \nat^{>0}$, let $a(n) = 0.$ Obviously, ${{a(n)}\over{n}} \to 0.$ If $p = 1,$ then for each $n \in \nat^{>0},$ let $a(n) = n.$ Obviously, ${{a(n)}\over{n}} \to 1.$ In both of these special cases, $a$ satisfies the required form and each $g_{ap}$ converges to the particular $p.$ \par\smallskip

Assume that $p \in (0,1)$ and the partition is made as described for $n \in \nat^{>0}.$  For $n = 1,$ let $a(1) = 0.$ Suppose that for $n >1$, $a(n)=k< n$ has been defined. Hence, ${{k}\over{n}}\leq p < 
{{k+1}\over{n}}$ and $\vert {{a(n)}\over{n}} - p \vert < {{1}\over {n}}.$ \par\smallskip

For the $n+1,$ partitioning yields  ${{k}\over{n+1}} < {{k}\over{n}}<{{k+1}\over{n+1}}<{{k+1}\over{n}}<
{{k+2}\over{n+1}},$ where we note that since $k < n,$ then $k+1 < n+1$, and it is possible that $k+2 = n+1.$ 
From the definition of the selection process, (1) if ${{k}\over{n+1}} < {{k}\over{n}}\leq p < {{k+1}\over{n+1}},$ then $a(n+1) = k<n.$ (2) If ${{k+1}\over{n+1}} \leq p < {{k+2}\over{n+1}},$ then $a(n+1) = k+1<n+1.$ In both cases, $\vert {{a(n+1)}\over{n+1}} -p \vert \leq {{1}\over {n+1}}$. Hence, by induction for all, $n \in \nat^{>0}$, if $a(1) = 0, $ then $a(n) = k, \  k< n,$ and $a(n+1) = k$ or $a(n+1) = k+1,$ and $\vert g_{ap}(n) -p\vert \leq 1/n.$ Thus, $a$ has the required form and ${{a(n)}\over{n}} \to p.$ These results show that each $A_p$ is nonempty. (A modification of the above process where you let $a(1) = 1$ will lead to the same conclusions.) \par\medskip

Once the sequence, in the above paragraph, is obtained, then denumerably many different sequences of this type can be defined that converge to the same $p \in [0,1].$ For the case where $p = 0,$ simply construct a sequence for each $m >1$ by letting $a(1) =0, a(n+1) = a(n) +1,$ for $1\leq n \leq m$, and then let $a(n+1) = a(n),\ n > m.$ Then
 ${{a(n)}\over{n}} \to 0.$ For the case where $p = 1,$ simply consider the sequence for each $m>1$, $a(n) = 0, \ 0 \leq n \leq m,$ and $a(n+1) = a(n) +1$ for each $n > m.$ Then ${{a(n)}\over{n}} \to 1.$\par\smallskip

Informally, for the case where $p \in (0,1)$, consider any member of $a \in A$ such that ${{a(n)}\over{n}} \to p.$  This sequence contains finite sets of consecutive members where the numerators are repeated. It contains denumerably many of these and the numerator numbers are strictly increasing for each collection of repeated members. For assume not. Then, from the definition, there exists some $k,j \in \nat^{>0}$ such that $g_{ap}(j) = {{k}\over{j}}$ and $g_{ap}(n) = {{k+n}\over{j+n}}$ for each $n >j.$ Since such a $g_{ap} \to 1,$ we have a contradiction. \par\smallskip

Take any one of these nonempty finite yet repeated collections of numerator numbers for $a$. Start at any one, and work backwards subtracting 1 from each of the previous numerators until you arrive at a zero and continue, if necessary, the remaining numerators as 0. One obtains, in each case, a $a\in A$ and ${{a(n)}\over{n}} \to p.$ There will be denumerably many different ones. \par\smallskip 

Suppose that one considers $a\in A$, where ${{a(n)}\over{n}} \to p \not= 0.$ Restrict $a$ to $n \leq m$ where, say, $m > 2.$ Then consider the sequence $a'$ such that $a'(k) = a(k),\ k \leq m$ and $a'(k)  =a(m), \ k> m.$ Clearly, ${{a'(n)}\over{n}} \to 0.$ Assume that this sequence comes from empirical evidence for a large number of trials less than $m$, where one is interested in successes (events occur). Then the known portion of the sequence would pass every statistical test with an appropriate confidence that ${{a'(n)}\over{n}} \to p.$ However, the actually behavior does not  follow the required convergence pattern after $m$ trials. Thus, operationally, it is not possible to establish by any mathematical test that a sequence of the required form actually converges, with any level of confidence, to $p.$ This can be extended to the notion of distributions as well. The probabilistic behavior for the occurrence of any such events is an assumption that cannot, in practice, be established formally. \par\smallskip

It is easy to construct members of $A$ that do not converge to any $p \in [0,1]$ (i.e. they contain subsequences that converge to different values). For example, start with ${{0}\over{1}}$. Now you increase the numerator number by 1 until you get ${{1}\over{2}}.$ Then  you repeat the numerator numbers until you get 
${{1}\over{3}}.$ Then increase each numerator by 1 until you get ${{1}\over{2}}$, etc. As an example, consider 
 ${{0}\over{1}},\ {{1}\over{2}},\ {{1}\over{3}},\ {{2}\over{4}},\ {{2}\over{5}},\ {{2}\over{6}},\ {{ 3}\over{ 7}},\ {{ 4}\over{ 8}},\ {{ 4}\over{ 9}},\ {{ 4}\over{10 }},\ {{ 4}\over{11 }},\ {{4 }\over{12 }},
\ {{ 5}\over{ 13}},\ {{6 }\over{ 14}},\ {{7 }\over{15 }},\ {{8 }\over{16 }},\ {{8 }\over{17 }},\ {{ 8}\over{18 }},\ {{8 }\over{19 }},\ {{8 }\over{20}},\ {{8 }\over{21 }},\ {{8 }\over{22 }},\ {{8 }\over{23 }},\break {{8 }\over{24 }},\ldots .$ The only thing one needs to do is to show that the points at which you alter the numerators or repeat the numbers will always occur after a finite number of steps.  Hence, $A'$ is also nonempty. Although such sequences are ``designed,'' it is often claimed that this type of cumulative event sequence can occur if physical-system behaved is purely random in character. However, note that Theorem 2.1 can be easily modified to show the existence of an ultralogic that generates any member of $A'$ if such a sequence does, indeed, model physical-system behavior. So, such members of $A'$ that correspond to certain aspects of how a physical-system developments, such as the notion of ``random'' fluctuations, can still be considered as designed by described algorithms.  \par\smallskip

The assumption that behavior is objectively probabilistic in character must come from some other source. However, this is also the case with all physical-system behavior that has some describable types of uniform behavior. Many accept the notion of the uniformity of nature. This uniformity can include not being able to predict via any human means the occurrence of an event or even exactly the composition of an event under describable conditions.  Nature would be uniform in that this would be the case under the specific conditions described. Again, the occurrence of such events can be considered as modeled by members of $A'$  and that they are designed. Further, hyperfinite choice would yield the actual event description. Thus, both conditions if they do, indeed, occur can still be the products of design.\par\smallskip  

(5/30/2013) The original paper was written prior to establishing the equivalence of finite consequence operators and logic-systems. Let the T portion of the language L include enough symbols to yield the set-theoretic representation for ordered pairs. Further, the tick marks can be replaced with corresponding natural number symbols. Indeed, the events E or E' need not cary this type of additional identifier. Although technically not necessary, T can include symbols for the simple ordering and the like. Of course, other natural number symbols and there first order properties are part of the formal standard structure employed and, as usually, are members of the meta-language. \par\smallskip.

It appears that the set $C_{ap}$ of consequence operators, can be replaced with the set $C_{ap}'=\{\r C(\{(j,M(P_{na(n)}(j)))\},\{\r G\})\mid 1\leq j \leq n\}.$ This yields the subtle consequence operator $P_p' \in \Hyper {\b H'}.$ This *consequence operator applied to $\{\b G\}$ yields the hyperfinite sequence 
$$S'=\{(1,\kern-.3em\hyper {M}(P_{\nu\kern-.3em\hyper {a(\nu)}}(1))), \ldots,(j,\kern-.3em\hyper {M}(P_{\nu\kern-.3em\hyper {a}(\nu)}(j))),\ldots, (\nu,\kern-.3em\hyper {M}(P_{\nu\kern-.3em\hyper {a}(\nu)}(\nu)))\}.$$ \par\smallskip

The hyperfinite sequence $S'$ forms a binary hyperfinite logic-system. When the rules of inference concept in [7] and the algorithm ``A'' for their use are formally expressed and  embedded into the above superstructure, this yields the formal hyper-algorithm $\hyper {\b A}$ and, when applied to each $j$ such that $1 \leq j\leq \nu$, this hyper-rationally yields an $\b E$ or an $\b E'$. As usually, each $\b E$ or $\b E'$ directly corresponds to the occurrence or non-occurences of a specific event. \par\smallskip

Thus, application of $P_p$ yields a coherent collection of events $\b E$ and $\b E'$ and indicates that they are each logically related via $P_p.$ Then application of $P_p'$ yields by hyper-deduction the actual occurrences. The order for results \b E or \b E' is *rationally designed via $P_p'.$ Relative to emerging properties, these subtle consequence operators demonstrate how a general collection of probabilistically guided physical events yields a  probability statement and maintains a *rational order even though members of the entire collection of such events appear to be independent one-from-another. \medskip

\centerline{\bf References}\medskip

\id{[1]} R. A. Herrmann, {\it Nonstandard Consequence Operators,} Kobe J. Math. {\bf 4}(1987), 1-14.
http://www.arXiv.org/abs/math.LO/9911204

\id{[2]} A. Tarski, {\it Logic, Semantics, Metamathematics; papers from} 1923 - 1938, Clarendon Press, Oxford, 1956.

\id{[3]} R. A. Herrmann, {\it The Theory of Ultralogics,} (1993),\hfil\break http://www.arXiv.org/abs/math.GM/9903081,  http://www.arXiv.org/abs/math.GM/9903082

\id{[4]} S. Kleene, {\it Mathematical Logic,} John Wiley \& Son, New York, 1967.
\id{[5]} R. A. Herrmann, {\it General logic-systems and consequence operators},\hfil\break http://arxiv.org/abs/math.GM/0512559
\id{[6]} B. S. DeWitt, {\it Quantum mechanics and reality}, Physics Today, 23(September 1970), 30-35.

\id{[7]} R. A. Herrmann, {\it Hyperfinite and Standard Unifications for Physical Theories,} Intern. J. Math. \& Math. Sci., 28(2)(2001), 93-102.\hfil\break http://www.arxiv.org/abs/physics/0105012\par

\end